\documentclass[referee,oldversion]{aa}
\usepackage{float}
\usepackage{graphicx}
\def\H0 {$H_{\rm o}$}

\def\CH3C2H {\hbox{${\rm CH}_3{\rm C}_2{\rm H}$}} 

\def\ffas {\hbox{$\,.\!\!^{\prime\prime}$}}
\def\ffcirc {\hbox{$\,.\!\!^{\circ}$}}

\def \ga{\mathrel{\mathchoice   {\vcenter{\offinterlineskip\halign{\hfil
$\displaystyle##$\hfil\cr>\cr\sim\cr}}}
{\vcenter{\offinterlineskip\halign{\hfil$\textstyle##$\hfil\cr
>\cr\sim\cr}}}
{\vcenter{\offinterlineskip\halign{\hfil$\scriptstyle##$\hfil\cr
>\cr\sim\cr}}}
{\vcenter{\offinterlineskip\halign{\hfil$\scriptscriptstyle##$\hfil\cr
>\cr\sim\cr}}}}}

\begin{document}

\title{DGSAT: Dwarf Galaxy Survey with Amateur Telescopes}
\subtitle{II. A catalogue of isolated nearby edge-on disk galaxies 
              and the discovery of new low surface brightness 
              systems}

\author{C. Henkel\inst{1,2} 
      \and 
       B. Javanmardi\inst{3} 
      \and
       D. Mart\'{\i}nez-Delgado\inst{4}
      \and
       P. Kroupa\inst{5,6}
      \and
       K. Teuwen\inst{7}} 

\offprints{C. Henkel, \email{chenkel@mpifr-bonn.mpg.de}}

\institute{
  Max-Planck-Institut f{\"u}r Radioastronomie, Auf dem H{\"u}gel 69, 53121 Bonn, Germany
 \and 
  Astronomy Department, Faculty of Science, King Abdulaziz University, P.O. Box 80203, 
  Jeddah 21589, Saudi Arabia
 \and
  Argelander Institut f{\"u}r Astronomie, Universit{\"a}t Bonn, Auf dem H{\"u}gel 71, 53121 Bonn, Germany
 \and
  Astronomisches Rechen-Institut, Zentrum f{\"u}r Astronomie, Universit{\"a}t Heidelberg, 
  M{\"o}nchhofstr. 12--14, 69120 Heidelberg, Germany
 \and
  Helmholtz Institut f{\"u}r Strahlen- und Kernphysik (HISKP), Universit{\"a}t Bonn, Nussallee 14--16, 
  D-53121 Bonn, Germany
 \and
  Charles University, Faculty of Mathematics and Physics, Astronomical Institute, 
  V Hole\v{s}ovi\v{c}k\'ach 2, CZ-18000 Praha 8, Czech Republic
 \and
  Remote Observatories Southern Alps, Verclause, France
}

\date{Received date ; accepted date}
 
\abstract
{
The connection between the bulge mass or bulge luminosity in disk galaxies and the 
number, spatial and phase space distribution of associated dwarf galaxies is a 
discriminator between cosmological simulations related to galaxy formation in cold dark 
matter and generalised gravity models. Here, a nearby sample of isolated Milky Way-class 
edge-on galaxies is introduced, to facilitate observational campaigns to detect the 
associated families of dwarf galaxies at low surface brightness. Three galaxy pairs with 
at least one of the targets being edge-on are also introduced. Approximately 60\% of the 
catalogued isolated galaxies contain bulges of different size, while the remaining objects 
appear to be bulgeless. Deep images of NGC~3669 (small bulge, with NGC~3625 at the edge of 
the image) and NGC~7814 (prominent bulge), obtained with a 0.4-m aperture, are also presented, 
resulting in the discovery of two new dwarf galaxy candidates, NGC~3669--DGSAT--3 and 
NGC~7814--DGSAT--7. Eleven additional low surface brightness galaxies are identified, 
previously notified with low quality measurement flags in the Sloan Digital Sky Survey 
(SDSS). Integrated magnitudes, surface brightnesses, effective radii, Sersic indices, 
axis ratios, and projected distances to their putative major hosts are displayed. At 
least one of the galaxies, NGC~3625--DGSAT--4, belongs with a surface brightness of  
$\mu_{\rm r}$ $\approx$ 26\,mag\,arcsec$^{-2}$ and effective radius $>$1.5\,kpc to the 
class of ultra-diffuse galaxies (UDGs). NGC~3669--DGSAT--3, the galaxy with the lowest 
surface brightness in our sample, may also be an UDG.
} 


\titlerunning{An edge-on sample of nearby galaxies}

\authorrunning{Henkel, C. et al.}

\maketitle

\section{Introduction}

The framework of the standard model of cosmology (SMoC) explains the striking difference 
between baryonic and apparent dynamical mass of galaxies (e.g. Lelli et al. 2016)
by introducing dark matter. According to this model, galaxies form in a hierarchical 
process with smaller dark matter halos merging to form more massive ones (e.g., Del 
Popolo 2014). While encounters between different halos may initially be hyperbolic, 
dynamical friction and dissipated energy leads to efficient coagulation. The massive 
dark halos of major host galaxies formed in this way do not only allow for high rotation 
velocities at large galactocentric radii but also contain a large number of subhalos, 
among them a notable fraction of dwarf galaxies. In view of the stochastic nature of 
galaxy encounters (but see Buck et al. 2015), individual infall histories, and the 
large number of objects involved, the spatial distribution of satellite galaxies should 
be spheroidal around the dominant host. And their number increases monotonically with 
circular velocity $V_{\rm rot}$ (and thus mass) of the host dark matter halo (Moore 
et al. 1999; Kroupa et al. 2010; Klypin et al. 2011; Ishiyama et al. 2013, Kroupa 2015).

The alternative approach, also successfully modelling orbital motions in galaxies, 
does not modify the amount of matter by introducing a dominant dark component, but -
based on baryonic matter alone - generalises instead gravity by introducing a scale 
invariant term when acceleration is below Milgrom's constant (a$_0$ $\approx$ 
3.8\,pc\,Myr$^{-2}$; Milgrom 1983a,b, 2009; Famaey \& McGaugh 2012; see also Moffat 2006). 
A significant fraction of dwarf galaxies may then be of tidal origin, their formation
being caused by hyperbolic encounters of larger galaxies (e.g. Yang et al. 2014). 
Such tidal dwarfs, if formed together in one encounter, should occupy a well defined, 
distinct region in six-dimensional phase space. 

It is of utmost importance to differentiate between these two basic concepts, related 
to galaxy evolution and the nature of the gravitational potential. This can be done 
by observationally testing predictions related to small-scale structure, for example,
by referring to the observable number of dwarf galaxies, their spatial distribution, 
and their distribution in phase space. For the standard model, the Local Group contains 
too few dwarf galaxies in comparison to the predicted number in the SMoC (e.g. Klypin 
et al. 1999; Simon \& Geha 2007) and dwarf galaxies are far from being distibuted 
isotropically around the Group's dominant members, the Andromeda galaxy (M~31) and 
the Milky Way. Instead, they show a highly flattened distribution around each of the 
galaxies, forming a Vast Polar Structure (VPOS) around the Milky Way and an edge-on 
viewed plain around M~31. In addition, isolated dwarf galaxies of the Local Group 
may be confined to two additional narrow planes (e.g. Kroupa et al. 2005; Ibata et 
al. 2013; Pawlowski et al. 2013, 2015; Pawlowski 2016, and, for attempts of simulation, 
e.g., Ahmed et al. 2017).

While this clearly disagrees with predictions of the $\Lambda$ cold dark matter 
($\Lambda$CDM) standard model, these data only shed light onto a single group of 
galaxies. Whether our Local Group is an extreme outlier or whether the dwarf galaxy
distribution of the Local Group is characterising the generic properties of the majority 
of groups in the local Universe is still an open question. Also, it is important to 
quantify the properties of satellite systems in the non-Newtonian framework. If these
are tidal dwarfs, then they further constrain the statistics of galaxy encounters and 
thus cosmological models. Less sensitive and less complete studies of nearby galaxies 
outside the Local Group (LG) also suggest highly flattened distributions, such as those 
around M~81, NGC3109, Cen~A, and M~83 (Bellazzini et al. 2013; Chiboucas et al. 2013; 
Ibata et al. 2014; M{\"u}ller et al. 2015, 2016, 2017b; Tully et al. 2015), but -- 
given the deep implications for fundamental physics and recent less conclusive results 
(e.g. Blauensteiner et al. 2017; M{\"u}ller et al. 2017a) -- more observations are 
mandatory to obtain a clearer picture. 

With this contribution we provide a catalogue of nearby disk galaxies, which are good 
candidates for probing the number and distribution of their satellites. Detailed
aspects of the motivation are outlined in Sect.~2, source selection criteria are 
discussed in Sect.~3, and the sample is introduced in Sect.~4. We also present
deep images from two of the sample galaxies in Sect.~5, followed by a short summary 
in Sect.~6.

\begin{figure*}
\vspace{0.0cm}
\centering
\hspace{-0.0cm}
\resizebox{15.5cm}{!}{\rotatebox[origin=br]{0.0}{\includegraphics{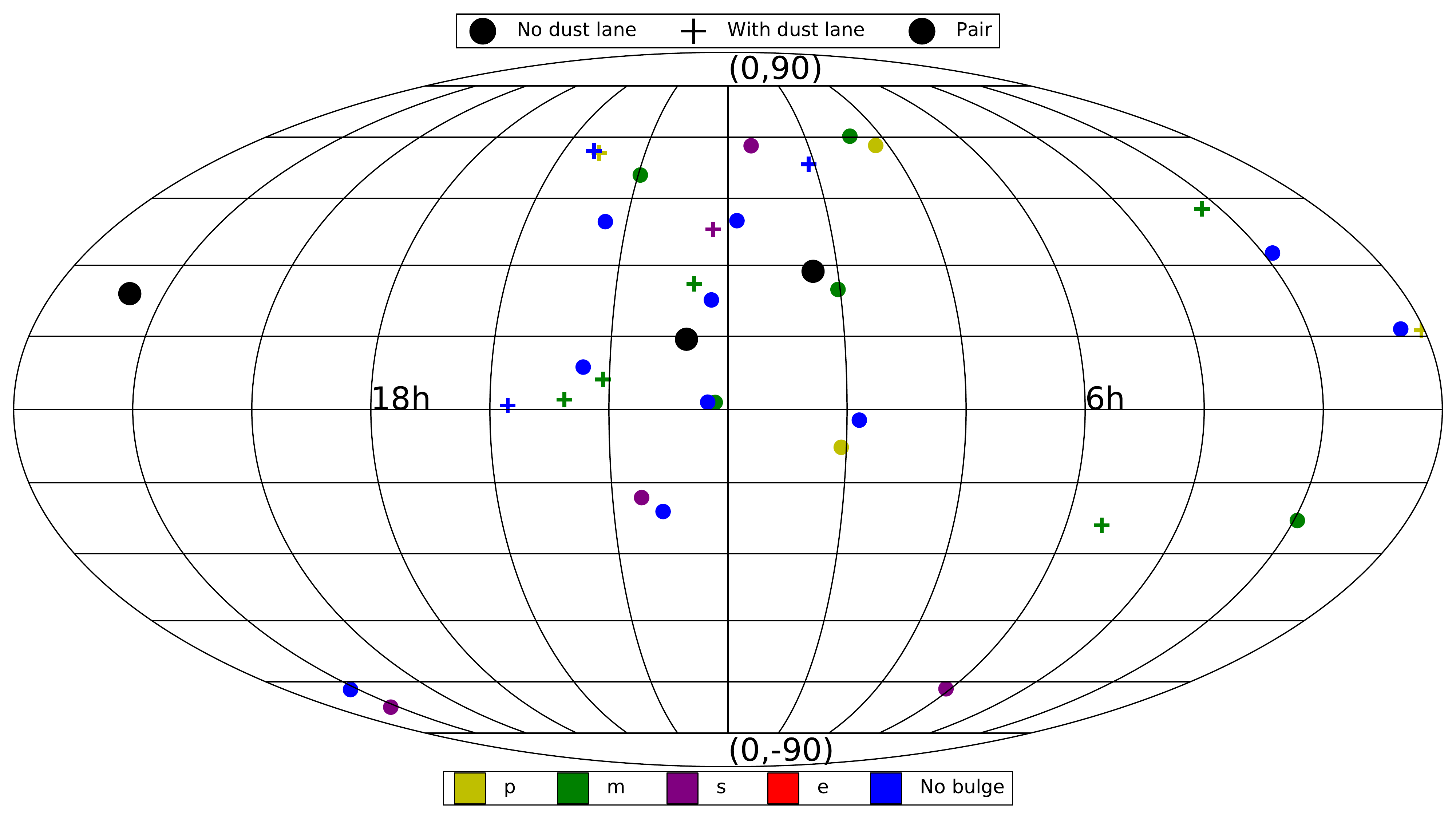}}}
\vspace{0.0cm}
\caption{Distribution of our sample of isolated galaxies (see Table~\ref{tab:singles}) 
and galaxy pairs (Table~\ref{tab:pairs}) in $J$2000 equatorial coordinates on the sky. 
Pairs are indicated by an outer circle while the inner symbol refers to their edge-on 
member. Bulge sizes (see Sect.\,4) and presence of equatorial dust lanes are also indicated.}
\label{allsky}
\end{figure*}

\section{Motivation}

A significant difference between the cosmological scenarios outlined above is related
to the formation of bulges. If structure formation is governed by dark matter, a large 
number of galaxy mergers should possibly lead to the early formation of bulges in each 
major galaxy located at the centre of a massive dark matter halo, with bulge mass 
related to the mass of the dark matter halo (e.g. Weinzirl et al. 2009; Kormendy et al.
2010; but see also Avila-Reese et al. 2014). Galaxy encounters in generalised gravity 
models (e.g. Renaud et al. 2016) should be rarer and not all of these should lead to 
mergers, since an absence of dark matter leads to a lesser amount of dynamical friction, 
allowing for a larger number of hyperbolic flybys. As a consequence, there may be more 
major galaxies with masses similar to the Milky Way, which do not show a bulge at all, 
consistent with observations (Fisher \& Drory 2011; Kroupa 2015) and with the gradual 
decrease of the fraction of bulgeless sources in the full sample of disk galaxies during 
the last $\approx$7.5 $\times$ 10$^9$\,yr (Sachdeva 2013). Such bulgeless galaxies may 
never have undergone a significant interaction, leading to dwarf galaxy populations devoid 
of tidal ones or, referring to our Local Group where most dwarfs appear to be tidal (e.g. 
Pawlowski et  al. 2015), to a few or even a total absence of dwarf galaxies. On the other 
hand, those galaxies having undergone significant encounters should show a bulge, accompanied 
by at least one plane of tidal dwarfs, possibly rotating in the same sense. 

As a consequence, with generalised gravity, dwarf galaxy populations of massive hosts 
with and without bulges should be very different. In the standard dark matter based
model, even bulgeless massive disk galaxies with high rotation velocity ($V_{\rm rot}$) 
should have many satellites, while in generalised gravity models the number of dwarfs 
should be positively correlated with bulge mass.  

Taking our Local Group as a template, where Andromeda (M~31), the Milky Way and 
the Triangulum galaxy (M~33) are the most prominent members, we find (Kroupa et 
al. 2010, their fig.~3) a rising number of dwarf spheroidal satellite galaxies
(dSph) with bulge mass. The correlation is consistent with a linear one, with M~33 
(no bulge) hosting none and M~31 (most massive bulge) being surrounded by the largest 
number of dSphs. However, as already mentioned above with respect to the Local Group, 
these results are suggestive but can still be challenged on statistical grounds (see, 
e.g., Metz et al. 2009 and Libeskind et al. 2009, 2015). Furthermore, one may ask 
whether M~33 can be termed an independent galaxy or merely a large satellite of the 
Andromeda galaxy. More recently, L{\'o}pez-Corredoira \& Kroupa (2016) found evidence 
for a positive correlation between the bulge/disk ratio of a major galaxy of given luminosity 
(from the Sloan Digital Sky Survey) and the number of associated tidal dwarf galaxies 
(from the catalogue of Kaviraj et al. 2012). 

If the number of dwarf galaxies is found not to be related to bulge properties but 
instead to $V_{\rm rot}$, this would weaken the argument that most dwarfs are of 
tidal origin. In any case, if we really want to provide basic constraints for 
cosmological models as well as the nature of gravity, the correlation between 
bulge mass and dwarf galaxy population has to be studied in much greater detail, 
requiring deep state of the art observations of a large number of galaxies,
including a reliable determination of their distances.

\begin{table*}
\caption[]{The sample of nearby edge-on galaxies as targets of dedicated searches for associated dwarf galaxies.}
\begin{flushleft}
\begin{tabular}{lccccrccl}
\hline
Name           &$\alpha_{J2000}$&$\delta_{J2000}$& $l^{\rm II}$ & $b^{\rm II}$  &   $V$        &    $D$        &log\,$L_{\rm B}$&  Bulge      \\
               &                &                &              &               &(km\,s$^{-1}$)&   (Mpc)       & (L$_{\odot}$)  &             \\
\hline 
               &                                 &              &               &              &               &                &             \\
NGC~7814       & 00 03 14.89    &+16 08 43.5     &     106.4    &  \,--45.2     &    1050      &  16$\pm$3     &     10.2       & p +         \\
NGC~100        & 00 24 02.98    &+16 29 13.6     &     113.5    &  \,--45.9     &     850      &  17$\pm$3     &      9.1       & --          \\
UGC~1281       & 01 49 31.61    &+32 35 19.5     &     136.9    &  \,--28.7     &     150      &   6$\pm$1     &      8.6       & --          \\
ESO~477--G016  & 01 56 15.99    & \,--22 54 04.0 &     200.4    &  \,--74.7     &    1825      &  36$\pm$5     &     10.1       & m           \\
NGC~891        & 02 22 33.41    &+42 20 56.9     &     140.4    &  \,--17.4     &     525      &  10$\pm$1     &     10.4       & m +         \\
NGC~1886       & 05 21 48.15    &\,--23 48 36.5  &     226.3    &  \,--29.6     &    1750      &  31$\pm$5     &      9.8       & m +         \\
ESO~121--G006  & 06 07 29.86    &\,--61 48 27.3  &     271.0    &  \,--28.8     &    1200      &  20$\pm$3     &      9.5       & s           \\
NGC~2549       & 08 18 58.35    &+57 48 11.0     &     159.7    &  +34.2        &    1075      &  17$\pm$7     &      9.8       & p           \\
NGC~2654       & 08 49 11.83    &+60 13 16.2     &     156.1    &  +37.8        &    1350      &  28$\pm$4     &     10.0       & m           \\
UGC~5245       & 09 47 34.49    &\,--02 01 57.1  &     238.8    &  +37.0        &    1400      &  21$\pm$4     &      9.0       & --          \\
               &                &                &              &               &              &               &                &             \\
NGC~3098       & 10 02 16.69    &+24 42 39.9     &     206.8    &  +52.1        &    1400      &  20$\pm$6     &      9.7       & m           \\
NGC~3115       & 10 05 13.98    &\,--07 43 06.9  &     247.8    &  +30.8        &     675      &  10$\pm$2     &     10.3       & p           \\
UGC~5459       & 10 08 10.07    &+53 05 00.5     &     160.8    &  +50.3        &    1100      &  22$\pm$4     &     10.1       & -- +        \\
NGC~3669       & 11 25 26.81    &+57 43 16.2     &     143.3    &  +55.9        &    1940      &  43$\pm$23    &     10.2       & s           \\
UGC~6792       & 11 49 23.30    &+39 46 17.4     &     164.5    &  +72.0        &     850      &  20$\pm$3     &      9.0       & --          \\
NGC~4179       & 12 12 52.11    &+01 17 58.9     &     281.6    &  +62.6        &    1250      &  18$\pm$4     &      9.9       & m           \\
UGC~7321       & 12 17 34.00    &+22 32 24.5     &     241.9    &  +81.1        &     400      &  18$\pm$6     &      9.1       & --          \\
NGC~4244       & 12 17 29.66    &+37 48 25.6     &     154.6    &  +77.2        &     250      &  04$\pm$2     &      9.7       & s +         \\
UGC~7394       & 12 20 27.66    &+01 28 11.1     &     285.5    &  +63.3        &    1600      &  32$\pm$3     &      9.2       & --          \\
NGC~4565       & 12 36 20.78    &+25 59 15.6     &     230.8    &  +86.4        &    1225      &  12$\pm$4     &     10.8       & m +         \\
	       &                &                &              &               &              &               &                &             \\
ESO~575--G061  & 13 08 15.60    &\,--21 00 08.0  &     308.2    &  +41.7        &    1650      &  29$\pm$5     &      9.4       & --          \\
NGC~5170       & 13 29 48.79    &\,--17 57 59.1  &     315.7    &  +44.0        &    1500      &  27$\pm$7     &     10.9       & s           \\
UGC~8880       & 13 57 11.48    &+50 26 08.6     &    \,\,99.1  &  +63.5        &    1850      &               &                & m           \\
NGC~5470       & 14 06 31.70    &+06 01 41.0     &     346.5    &  +62.4        &    1025      &  24$\pm$3     &      9.2       & m +         \\
UGC~9242       & 14 25 21.02    &+39 32 22.5     &    \,\,71.4  &  +66.9        &    1475      &  20$\pm$6     &      9.0       & --          \\
UGC~9249       & 14 26 59.78    &+08 41 01.0     &     358.2    &  +60.9        &    1375      &  19$\pm$3     &      8.9       & --          \\
NGC~5746       & 14 44 55.92    &+01 57 18.0     &     355.0    &  +53.0        &    1725      &  30$\pm$3     &     11.1       & m +         \\
NGC~5866       & 15 06 29.50    &+55 45 47.6     &    \,\,92.0  &  +52.5        &     750      &  12$\pm$4     &     10.1       & p +         \\
NGC~5907       & 15 15 53.77    &+56 19 43.6     &    \,\,91.6  &  +51.1        &     675      &  16$\pm$2     &     10.8       & -- +        \\
UGC~9977       & 15 41 59.54    &+00 42 46.1     &  \,\,\,\,7.3 &  +41.3        &    1925      &  29$\pm$3     &     10.3       & -- +        \\
               &                &                &              &               &              &               &                &             \\
ESO~146--G014  & 22 13 00.43    &\,--62 04 03.4  &     328.6    & \,--46.4      &    1700      &  21$\pm$2     &      9.4       & --          \\
IC~5176        & 22 14 55.93    &\,--66 50 57.9  &     323.0    & \,--43.7      &    1750      &  27$\pm$3     &     10.3       & s           \\
	       &                &                &              &               &              &               &                &             \\
\hline
\end{tabular}
\end{flushleft}
Col.\,(1): Source name \\
Cols.\,(2) -- (5): Equatorial and Galactic coordinates \\
Col.\,(6): Systemic velocity from NED \\
Col.\,(7): Here and elsewhere in the text, distances $D$ are average values from the literature given by NED. If there is no distance 
determination independent of radial velocity, no value is given. \\
Col.\,(8): The logarithm of the blue luminosity ($L_{\rm B}$ in solar units) was taken from the ``RC3'' of de Vaucouleurs et al. (1991), 
correcting for inclination, internal extinction, and redshift. Only for NGC~1886, ESO~121--G006, and ESO~311--G012 were luminosities 
calculated from RC3 uncorrected blue magnitudes. \\
Col.\,(9): -- indicates the apparent absence of a bulge; otherwise, with increasing dominance of a bulge, small = s, moderate = m, and 
prominent = p (see Sect.\,4). A `+' indicates that at least some traces of an equatorial dust lane are present. \\
\label{tab:singles}
\end{table*}

\begin{table*}
\caption[]{Nearby galaxy pairs}
\begin{flushleft}
\begin{tabular}{lccccccccl}
\hline
\multicolumn{2}{c}{Names}         & $\alpha_{J2000}$ & $\delta_{J2000}$ & $l^{\rm II}$ & $b^{\rm II}$ &     $V$      &   $D$         & log\,$L_{\rm B}$      & Bulge        \\
               &                  &                  &                  &              &              &(km\,s$^{-1}$)&   (Mpc)       &  (L$_{\odot}$)        &              \\
\hline 
               &                  &                  &                  &              &              &              &               &                       &              \\
NGC~3245A      & NGC~3245         & 10 27 01.13      & +28 38 21.6      &  201.6       &   +58.2      &    1325      & 25$\pm$3      &             9.8       & s            \\
NGC~4634       & NGC~4633         & 12 42 40.96      & +14 17 45.0      &  293.5       &   +77.0      &     300      & 21$\pm$2      &             9.4       & e +          \\
NGC~7332       & NGC~7339         & 22 37 24.54      & +23 47 54.0      &  \,\,87.4    & \,--29.7     &    1200      & 15$\pm$6      &             9.8       & p            \\
	       &                  &                  &                  &              &              &              &               &                       &              \\
\hline
\end{tabular}
\end{flushleft}
Cols.\,(1) and (2): Names of the edge-on galaxy and its companion. All following parameters are those of the edge-on galaxy. 
NGC~3245 is an SA0 galaxy that is not seen edge-on, NGC~4633 is a spiral at intermediate and NGC~7339 a spiral at high inclination. \\
Cols.\,(3) -- (6): Equatorial and Galactic coordinates \\
Col.\,(7): Systemic velocity from NED \\
Col.\,(8): Distances $D$ are average values from the literature given by NED. For NGC~4634 the distance was obtained from its companion 
NGC~4633. We note, however, that this distance estimate is only based on two independent estimates.  \\
Col.\,(9): The logarithm of the blue luminosity ($L_{\rm B}$ in solar units) was taken from ``RC3'' (de Vaucouleurs et al. 1991), 
correcting for inclination, internal extinction, and redshift. \\
Col.\,(10): small bulge = s, prominent bulge = p and boxy extended bulge = e.  +: at least some traces of an equatorial 
dust lane are present. \\
\label{tab:pairs}
\end{table*}

\section{Selection criteria to find suitable targets}

To achieve this goal it is useful to compile a catalogue of suitable host galaxies. 
The most straightforward way is to focus on Milky Way-class edge-on L$^*$ galaxies, 
where the presence or absence of a bulge is clearly seen and where a bulge mass 
can be derived. This is our {\it first criterion}. Such a sample with a single clear-cut 
viewing angle also allows for clear determinations of rotation curves. Furthermore, 
it will avoid pitfalls related to geometrical effects and will provide a detailed 
comparison not only between objects of different bulge mass but also between sources 
with similar bulge to disk masses or luminosities. This implies that also the scatter 
among objects with similar properties can be evaluated. Finally, the inclination of 
the potentially observed families of dwarf galaxies may reveal some scatter of 
angles with respect to the host's large scale plane, depending on the kind of 
interactions in the past. 

Focusing on the environment of edge-on galaxies will greatly facilitate the 
determination of angles between their disks and potential planes of dwarf 
galaxies. Targets of choice should thus be those spiral galaxies, where an edge-on 
view is guaranteed. Then, for example, in case of a polar structure seen edge-on, 
the satellites would form a line similar to that shown by Kroupa et al. (2005; 
their fig.~1) with significant shifts in radial velocity with respect to systemic 
at the outer edges ($V$\,sin\,$\theta$, with $V$ denoting the rotation velocity 
and $\theta$ representing the angle of the line from the centre of the galaxy 
to the dwarf with respect to our line of sight). In case such a structure is 
seen face-on, satellites should show a rather isotropic distribution, but with 
no significant changes in radial velocity because of orbital motions being 
mostly confined within the plane of the sky.  

Because extremely deep exposures are required to detect as many dwarfs as possible
(e.g. Merrit et al. 2014; Javanmardi et al. 2016), the field of view of optical 
telescopes is another important parameter. It should allow for direct imaging 
of the entire relevant volume surrounding a host galaxy. This is difficult to achieve. 
L$^*$ disk galaxies are characterised by  $M$ $\approx$ 10$^{12}$\,M$_{\odot}$ dark 
matter halos with virial radii $r_{\rm vir}$ $\approx$ 250\,kpc (e.g. Kravtsov 2013;
Taylor et al. 2016), which must have grown from a sequence of mergers if the Standard 
Model of Cosmology is the valid theory. In the case of Andromeda, Perseus I, also called 
Andromeda XXXIII, is located at a distance of 375$\pm$15\,kpc from its host (Martin et al. 
2013). Imaging projected radii out to 375\,kpc, for a 0$^{\circ}\!\!$.5 field of view (FoV), 
distances $D$ $\ga$ 85\,Mpc and for a 3$^{\circ}$ FoV, $D$ $\ga$ 15\,Mpc are required, 
corresponding to recession velocities of approximately $\ga$12000 and $\ga$2000\,km\,s$^{-1}$, 
respectively. For a more modest projected diameter of 250\,kpc, centred on a given 
host, 0$^{\circ}\!\!$.5 and 3$^{\circ}$ FoVs require $D$ $\ga$30 and $\ga$4\,Mpc, 
corresponding to recession velocities of $\ga$2000 and $\ga$300\,km\,s$^{-1}$,
respectively. Thus, field sizes well in excess of 1$^{\circ}$ would be desirable. 

Our {\it second criterion}, in clear opposition to the need for a coverage of
large linear scales, is the requirement that the targets have to be spatially 
resolved. The detection of faint satellites is already difficult in the Local Group 
and much larger distances would lead to an overly high number of undetectable 
low-luminosity satellites. Resolving the galaxies is essential to prove that they 
are galaxies beyond doubt and to determine their surface brightness, effective 
radius ($r_{\rm e}$), and magnitude. A characteristic $r_{\rm e}$ = 500\,pc (e.g.
Javanmardi et al. 2016) corresponds to 3$''$ at a distance of 30\,Mpc 
($\approx$2000\,km\,s$^{-1}$ recessional velocity; see also fig.~5 of Carrasco et al. 
2006)). This distance is close to the limit that could be reached by telescopes 
equipped with wide-field cameras and is, to provide a comparison, 2-5 times larger 
than the putative distances to the dwarf galaxy candidates identified by 
Merritt et al. (2014), Karachentsev et al. (2015), and Javanmardi et al. (2016).

The {\it third criterion}, again in opposition to the requirement of large linear 
scales to be covered simultaneously, is caused by field contamination due to 
unwanted foreground galaxies with their extended (on angular scales) foreground 
satellite systems. Thus again nearby galaxies are the targets of choice,
while background sources covering substantially smaller solid angles may be 
recognised by their redshift (in case they are major galaxies) or by other 
properties (we refer to the end of this section and Sect.\,5).

Finally, we have to avoid Galactic cirrus, which may inhibit a successful search 
for dwarfs often characterised by a low surface brightness. Therefore the galaxies 
should not be located close to the Galactic plane. This is the {\it fourth 
criterion}. 

In the following section, we provide a list of suitable nearby edge-on targets, which 
are devoid of any nearby companions of similar angular size. As a caveat, we must
note that although it is important to identify the dwarfs' association with 
their hosts beyond doubt, distances of such tiny galaxies would not be easy to 
obtain in a direct way (e.g. through brightest supergiant or tip of the 
red giant branch (TRGB) observations). Nevertheless, plots relating luminosity or 
surface brightness to effective radius or surface brightness to absolute magnitude 
(e.g. Javanmardi et al. 2016) can help to provide, beside a limited angular 
offset to the apparent host, significant first evidence for such an association
(we refer to Sect.\,5).

\section{The sample}

Among the sample of 1222 edge-on spiral galaxies in total from NED\footnote{The NASA/IPAC 
Extragalactic Database (NED) is operated by the Jet Propulsion Laboratory, California 
Institute of Technology, under contract with the National Aeronautics and Space 
Administration.}, 267 are close enough to match our criterion of $V$ $<$ 
2000\,km\,s$^{-1}$ (Sect.\,3), which allows us to spatially resolve the objects but
requires wide fields ($>$0.$\!\!^\circ$5) of view. In addition, among the 55 S0/a 
and 391 S0 galaxies, 12 and 89 are characterised by velocities $\leq$2000\,km\,s$^{-1}$, 
respectively. These 368 nearby sources were individually checked through images from 
NED and the SIMBAD Astronomical Database\footnote{This research has made use of the SIMBAD 
database, operated at CDS, Strasbourg, France}. In some cases ratings of ``very good'' 
were assigned, while the remaining ones were discarded. Galaxies showing significant 
distortions were not included. Galaxies not exactly seen edge-on were downgraded. 
A frequently arising question was, whether an extended bulge was existing or 
the galaxy was instead not quite edge-on. These galaxies, mostly suffering from a 
limited image quality, were also excluded. In this way, 21 isolated edge-on targets
could be identified.

\begin{figure*}
\begin{center}
\includegraphics[scale=0.8]{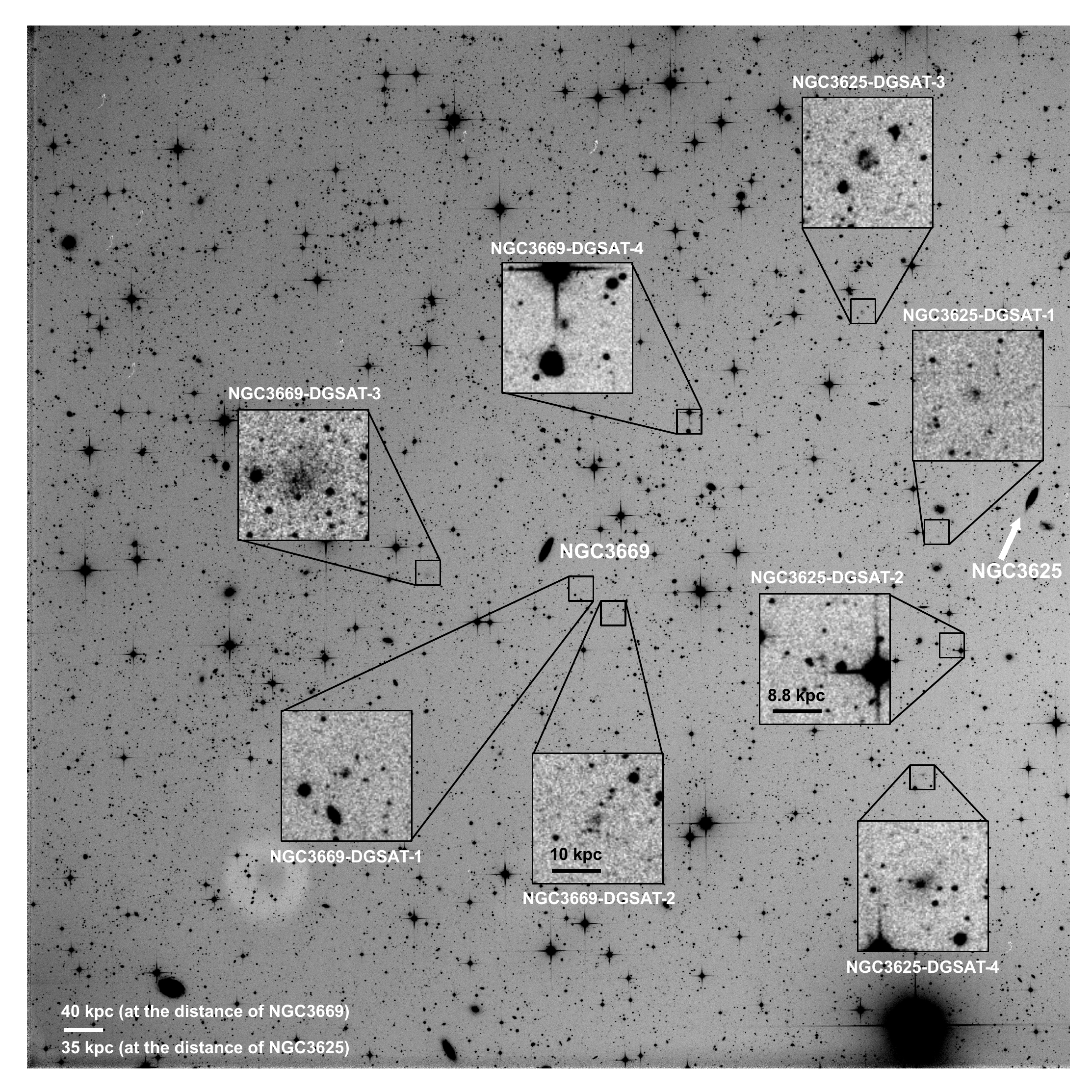}
\caption{A $1^\circ\!\!.4 \times 1^\circ\!\!.4$ image centred on NGC 3669. North is up 
and east is to the left. The zoomed-in squares of size 101$\times$101 pixels (125$''$$\times$125$''$)
highlight the LSB (Low Surface Brightness) galaxies in the image.
\label{fig:ngc3669}}  
\end{center}
\end{figure*}

\begin{figure*}
\begin{center}
\includegraphics[scale=0.79]{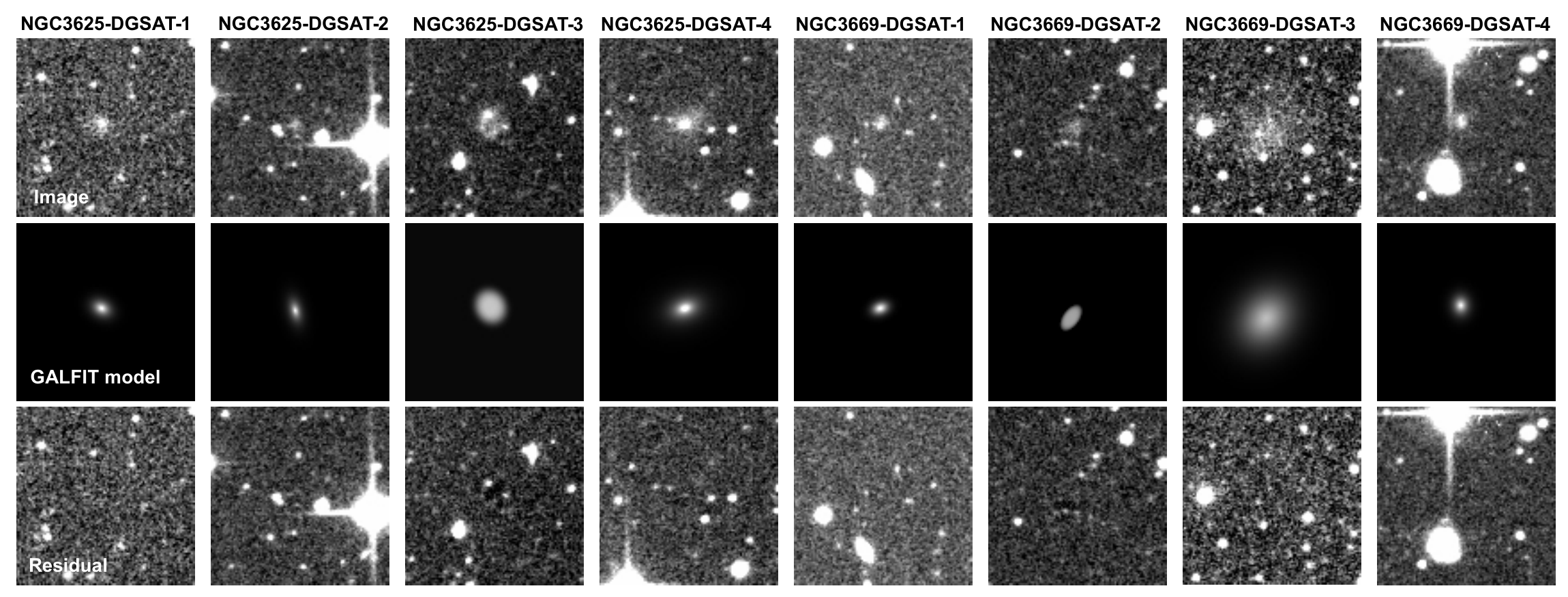}
\caption{Top row: dwarf galaxy candidates detected in our image; middle row: our GALFIT 
models (see Table \ref{tab:results} for the results of the modelling), and bottom row: 
residual images obtained by subtracting the model image (middle) from the main image (top). 
As the inserts in Fig.~\ref{fig:ngc3669}, the images have a size of 101$\times$101 pixels, 
corresponding to 125$''$$\times$125$''$. North is up and east is to the left.
\label{fig:3669_models}}
\end{center}
\end{figure*}

\begin{table*}
\begin{center}
\caption{LSB galaxies, their equatorial coordinates, RA and DEC ($J$2000), and their structural 
parameters evaluated by GALFIT modelling (see Peng et al. 2002 and Javanmardi et al. 2016); 
calibrated integrated magnitude, $r_{cal}$, surface brightness at effective radius, $\mu_{e}$, 
the Sersic index, $n$, the effective radius, $R_e$, and the axis ratio, $b/a$. 
\label{tab:results}}
\begin{tabular}{cccccccc}
 ID & $\alpha_{J2000}$ & $\delta_{J2000}$ & $r_{cal}$ {\scriptsize(mag)} & $\mu_{e}$ {\scriptsize(mag/arcsec$^2$)} & $n$ & $R_e$ {\scriptsize (arcsec)} & $b/a$\\
  \hline
  \hline
 {\small NGC3625-DGSAT-1} & 11 21 28.45  & +57 43 55.2  &  $19.9\pm0.1$ & $26.2\pm0.1$  & $0.83\pm0.08$  & $6.3\pm0.5$  & $0.78\pm0.04$ \\
 {\small NGC3625-DGSAT-2} & 11 21 22.91  & +57 34 50.1  &  $20.1\pm0.3$ & $26.7\pm0.3$  & $1.04\pm0.23$  & $8.5\pm1.9$  & $0.53\pm0.06$ \\
 {\small NGC3625-DGSAT-3} & 11 22 12.03  & +58 02 11.9  &  $18.8 \pm 0.1$ & $25.3 \pm 0.1$  & $0.26\pm0.02$  & $8.2 \pm 0.2$  & $0.87\pm0.02$ \\
 {\small NGC3625-DGSAT-4} & 11 21 40.79  & +57 24 37.0  &  $18.3 \pm 0.1$ & $26.1 \pm 0.1$  & $1.24\pm0.07$  & $12.0 \pm 0.8$  & $0.70\pm0.02$ \\
 \hline
 {\small NGC3669-DGSAT-1} & 11 25 04.39  & +57 39 53.6  &  $20.2 \pm 0.1$ & $25.8 \pm 0.1$  & $0.66\pm0.07$  & $5.1 \pm 0.4$  & $0.73\pm0.04$ \\
 {\small NGC3669-DGSAT-2} & 11 24 48.32  & +57 37 58.0  &  $19.8 \pm 0.1$ & $26.0 \pm 0.1$  & $0.19\pm0.03$  & $8.9 \pm 0.3$  & $0.56\pm0.02$ \\
 {\small NGC3669-DGSAT-3} & 11 26 38.78  & +57 41 19.1  &  $18.1 \pm 0.1$ & $26.7 \pm 0.1$  & $0.65\pm0.04$  & $18.6 \pm 0.9$  & $0.83\pm0.03$ \\
 {\small NGC3669-DGSAT-4} & 11 23 57.86  & +57 53 25.6  &  $19.3 \pm 0.1$ & $25.7 \pm 0.1$  & $0.83\pm0.05$  & $6.3 \pm 0.3$  & $0.82\pm0.03$ \\
 \hline
 {\small NGC7814-DGSAT-1} & 0 03 24.07  & +16 11 11.1  &  $17.7 \pm 0.1$ & $26.0 \pm 0.1$  & $0.59\pm0.03$  & $15.8 \pm 0.7$  & $0.86\pm0.03$ \\
 {\small NGC7814-DGSAT-2} & 0 03 06.93  & +16 18 30.8  &  $17.7 \pm 0.1$ & $26.7 \pm 0.1$  & $1.04\pm0.05$  & $29.4 \pm 1.5$  & $0.35\pm0.01$ \\
 {\small NGC7814-DGSAT-3} & 0 04 11.75  & +16 32 01.2  &  $18.1 \pm 0.1$ & $25.5 \pm 0.1$  & $0.78\pm0.03$  & $11.6 \pm 0.3$  & $0.62\pm0.01$ \\
 {\small NGC7814-DGSAT-4} & 0 02 48.89  & +16 35 50.9  &  $18.7 \pm 0.1$ & $25.8 \pm 0.1$  & $0.78\pm0.06$  & $8.5 \pm 0.5$  & $0.92\pm0.04$ \\
 {\small NGC7814-DGSAT-5} & 0 02 35.21  & +16 42 53.4  &  $20.6 \pm 0.1$ & $25.7 \pm 0.1$  & $0.39\pm0.07$  & $4.7 \pm 0.3$  & $0.62\pm0.04$ \\
 {\small NGC7814-DGSAT-6} & 0 00 27.94  & +16 11 00.6  &  $18.8 \pm 0.1$ & $24.9 \pm 0.1$  & $0.80\pm0.04$  & $5.9 \pm 0.2$  & $0.72\pm0.02$ \\
 {\small NGC7814-DGSAT-7} & 0 00 43.95  & +15 27 14.3  &  $18.5 \pm 0.1$ & $26.8 \pm 0.1$  & $1.14\pm0.11$  & $19.1 \pm 2.1$  & $0.49\pm0.02$ \\
\hline
\end{tabular}
\end{center}
\end{table*}

A second sample has been drawn from the $\approx$5750 sources presented by Bizyaev et al. 
(2014) in their catalogue of edge-on disk galaxies, based on the SLOAN Digital Sky Survey 
(SDSS). Limiting the sample again to $V$ $\leq$ 2000\,km\,s$^{-1}$, 171 sources were
extracted and individually checked. Using the same procedure as for the NED objects, 
22 were rated to be ``very good''. Not surpringly, there is some overlap with NED so 
that the number of additional sources is only 16. Two of them, even though showing 
radial velocities close to 0\,km\,s$^{-1}$, are so faint that bulge properties could
not be assigned. Excluding these two, which may show low intrinsic luminosity, that is
PGC~1653789 and SDSS~J081732.25+393629.0, then leads to 21+14 = 35 galaxies.

Also applying the fourth criterion (Sect.\,3) and requesting a sufficient angular 
distance from the Galactic plane, we find three sources with Galactic latitude
$|b^{\rm II}|$ $<$ 15$^{\circ}$. These were discarded, leaving a total of 32 targets. 
Among the latter, only NGC~891 is located at a $|b^{\rm II}|$ slightly below 20$^{\circ}$. 
For all others, $|b^{\rm II}|$ $>$ 28$^{\circ}$.

To summarise, the selected sub-sample is not free of observational bias, depending on the 
quality of the images and the brightness of the respective target. Thus, there may be 
suitable galaxies rejected by our harsh selection criteria. However, the goal of this 
article is not to present a complete sample but to guarantee that each selected galaxy 
is genuine. The best candidates are truly edge-on viewed bulgeless targets as well 
as galaxies with bulges, where either a dust lane partially obscuring the nuclear region 
clearly identifies the equatorial plane and/or where the outer disk is extremely thin, 
consistent with an edge-on view. Already a small number of such galaxies are sufficient 
to provide an answer to the questions raised in Sects.\,1--3. 

Table~\ref{tab:singles} presents the source list, consisting of 32 Milky Way-class 
objects with log\,($L_{\rm B}/L_{\odot}$) = 8.6 -- 11.1. Infrared luminosities $L_{\rm IR}$
obtained by fitting IRAS data and extrapolating to the full range of 8 -- 400\,$\mu$m
(for details, see Wouterloot \& Walmsley 1986) tend to be lower or close to the blue 
luminosities (exceptions: ESO~121--G006 and, see below, NGC~4634 with $L_{\rm IR}$/$L_{\rm B}$ 
$\approx$ 3). None of our objects is a vigorously active galaxy (with a recent starburst 
totally dominating $L_{\rm B}$), thus ensuring that our blue luminosities represent 
approximately the masses of the listed galaxies. Their distribution on the sky is shown 
in Fig.~\ref{allsky}. Of our sources 11 exhibit at least some trace of an equatorial dust 
lane, providing a particularly reliable confirmation of an edge-on view. Of the 32 sources, 
13 appear to be bulgeless. The remaining 19 targets have bulges of different prominence, 
ranging from small to prominent. The bulges are classified following their extent along 
the major axis of a given host relative to the total extent of the edge-on viewed disk 
and contain the following specifications: 

{\bf Prominent}: The bulge extends over 45\% -- 85\% of the length of the visible plane 
of the galaxy. 

{\bf Moderate}: The bulge covers 20\% -- $<$45\% of the visible length of the host.

{\bf Small}: An inconspicious bulge may be present.

Therefore, targets of all classes could be studied without contamination from other nearby 
sources. While the galaxies show no similarly sized nearby counterpart, the fields
are nevertheless often not completely empty. Therefore, details of the environment 
of individual edge-on targets are discussed, source by source, in the Appendix.  

While the presence of many Milky Way-sized galaxies in groups or clusters would lead
to galaxy distributions which are too complex to analyse, the presence of two relatively 
nearby L$^*$ objects may still not provide unsurmountable problems. Furthermore, the
Local Group is based on a pair of two such galaxies, the Milky Way and M~31. It is
by far the best studied galaxy group and naturally serves as THE template. Whether such 
pairs show a stochastic superposition of the dwarf galaxy populations expected for 
individual parent galaxies or whether the distributions are correlated in some way as 
proposed by Libeskind et al. (2016) on the basis of SDSS data remains an open question. 
Potentially interesting in this context are the pairs NGC~3245 -- NGC~3245A, NGC~4633 
-- NGC~4634, and NGC~7332 -- NGC~7339. Relevant data from the edge-on members of these 
pairs are presented in Table~\ref{tab:pairs} and further information is given, as for 
the isolated Milky Way-sized targets in the Appendix.

\begin{figure*}
\begin{center}
\includegraphics[scale=0.8]{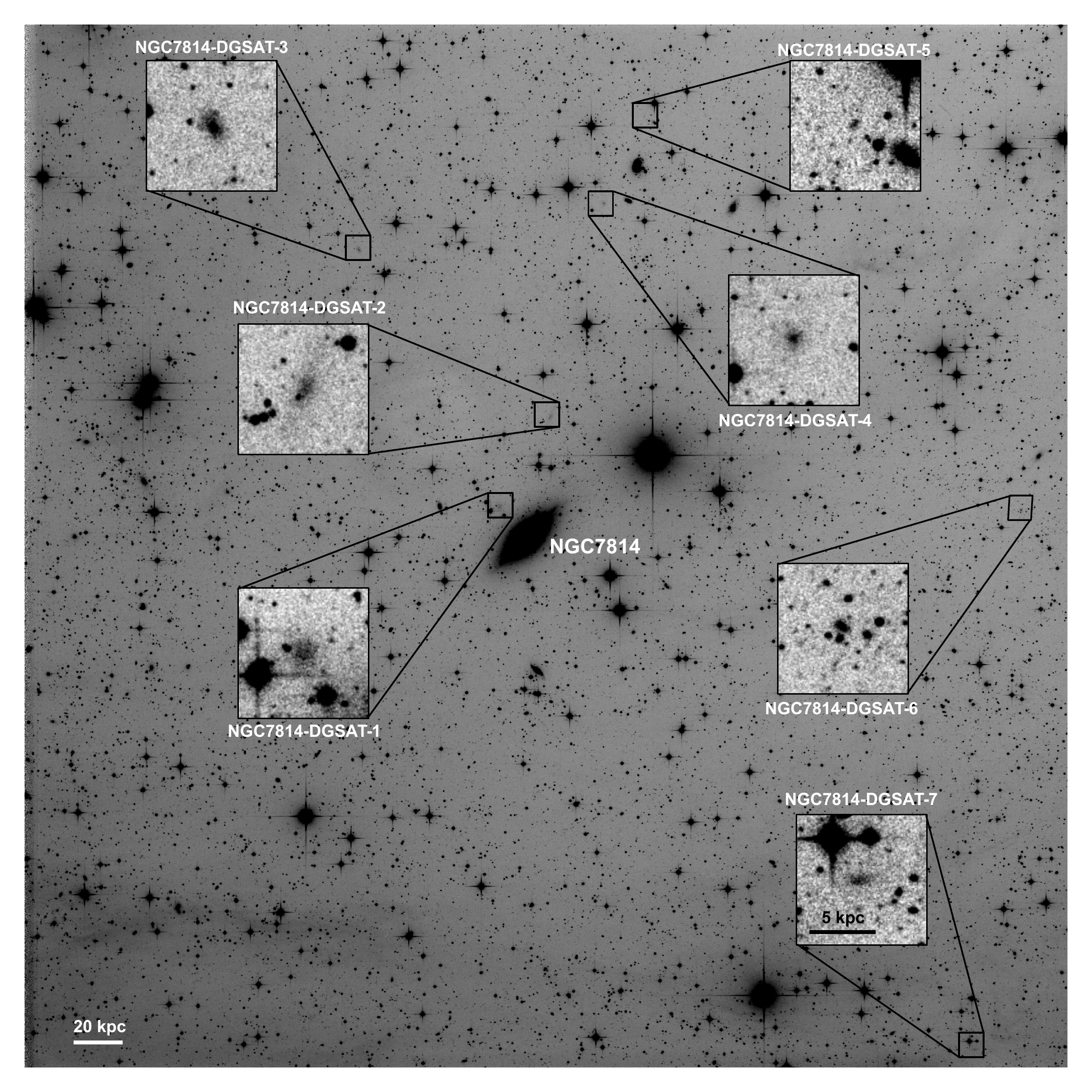}
\caption{The $1^\circ\!\!.4 \times 1^\circ\!\!.4$ field of NGC 7814. North is 
up and east is to the left. The zoomed-in squares highlight the LSB galaxies in 
the image. The angular size of the inserts is the same as in Fig.~\ref{fig:ngc3669}.
\label{fig:ngc7814}}
\end{center}
\end{figure*}

\begin{figure*}
\begin{center}
\includegraphics[scale=0.9]{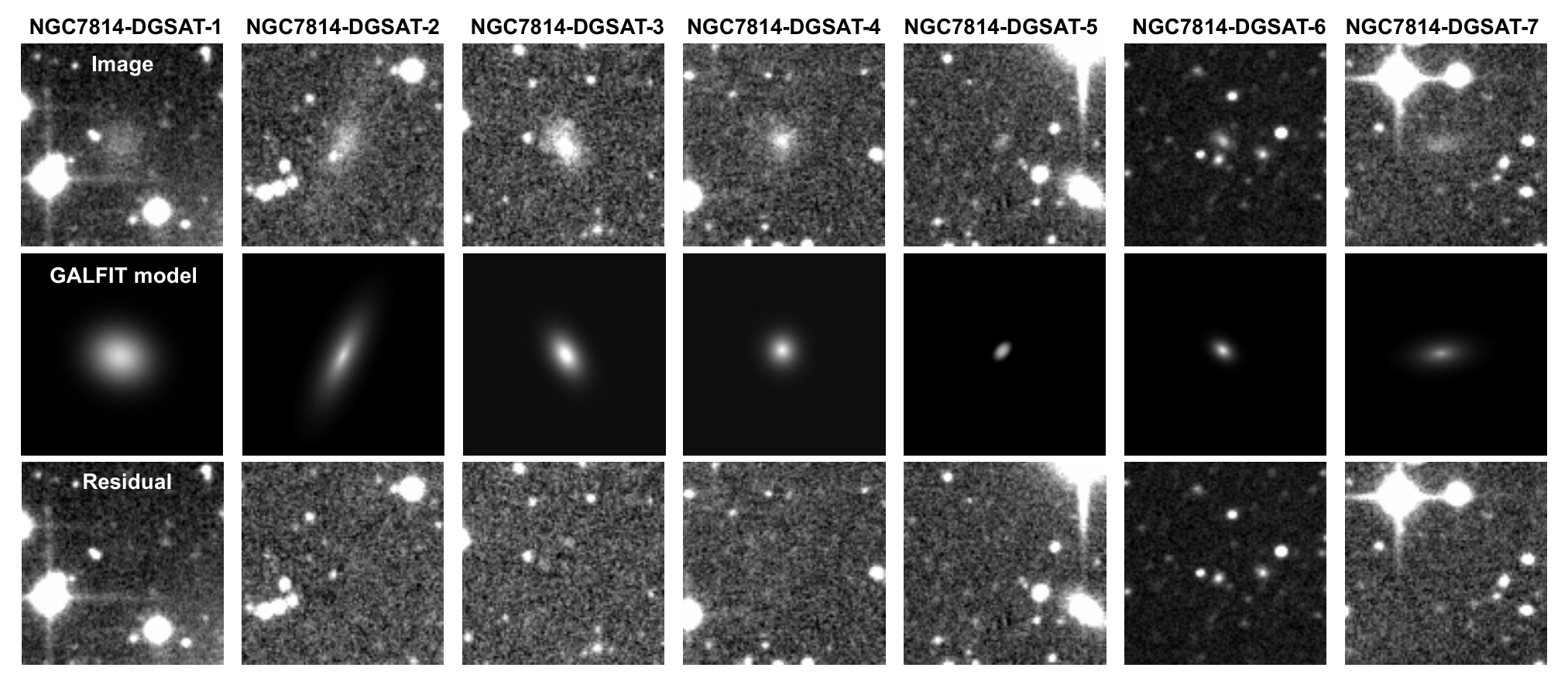}
\caption{Top row: dwarf galaxy candidates detected in our image; middle 
row: our GALFIT models (see Table \ref{tab:results} for the results of the modelling), 
and bottom row: residual images obtained by subtracting the model image (middle) 
from the main image (top). Orientation and angular size of the images are the same
as in Fig.~\ref{fig:3669_models}.
\label{fig:7814_models}}
\end{center}
\end{figure*}

\begin{table*}
\begin{center}
\caption{Physical properties of the LSB galaxies assuming that they are satellites 
of the nearby (in projection) massive galaxies. $R_e$ and $d_P$ are effective radius 
and projected distance of the dwarf candidate to the centre of the host galaxy, 
respectively. $M_r$ denotes the absolute magnitude in r-band and the last column 
provides the logarithm of the r-band luminosity in units of solar luminosities, 
L$_{\odot}$.  
\label{tab:results_phys}}
  \begin{tabular}{lllll}
 ID & $R_e$ {\scriptsize(pc)} & $d_P$ {\scriptsize(kpc)}& M$_r$ {\scriptsize(mag)}&$\log\left(\frac{L}{L_{\odot}}\right)_r$\\
  \hline
  \hline
 {\small NGC3625-DGSAT-1} & $1145 \pm 156$ & $ 89.6 \pm 9.8 $ & $-13.0 \pm 0.2$ & $ 7.1 \pm 0.1 $ \\
 {\small NGC3625-DGSAT-2} & $1536 \pm 380$ & $ 151.2 \pm 16.6 $ & $-12.8 \pm 0.4$ & $ 7.0 \pm 0.2 $ \\
 {\small NGC3625-DGSAT-3} & $1480 \pm 166$  & $ 220.6 \pm 24.2 $ & $-14.1 \pm 0.2$ & $ 7.5 \pm 0.1 $ \\
 {\small NGC3625-DGSAT-4} & $2175 \pm 275$ & $ 262.7 \pm 28.9 $ & $-14.5 \pm 0.2$ & $ 7.7 \pm 0.1 $ \\
 \hline
 {\small NGC3669-DGSAT-1} & $1064 \pm 572$ & $ 57.4 \pm 30.6 $ & $-12.7 \pm 1.2$ & $ 7.0 \pm 0.5 $ \\
 {\small NGC3669-DGSAT-2} & $1864 \pm 996$ & $ 93.0 \pm 49.5 $ & $-13.1 \pm 1.2$ & $ 7.1 \pm 0.5 $ \\
 {\small NGC3669-DGSAT-3} & $3912 \pm 2094$ & $ 123.1 \pm 65.6 $ & $-14.8 \pm 1.2$ & $ 7.8 \pm 0.5 $ \\
 {\small NGC3669-DGSAT-4} & $1325 \pm 709$ & $ 196.6 \pm 104.8 $ & $-13.6 \pm 1.2$ & $ 7.3 \pm 0.5 $ \\
  \hline
 {\small NGC7814-DGSAT-1} & $1231 \pm 269$ & $ 15.6 \pm 3.4 $ & $-13.2 \pm 0.5$ & $ 7.2 \pm 0.2 $ \\
 {\small NGC7814-DGSAT-2} & $2285 \pm 504$ & $ 46.7 \pm 10.0 $ & $-13.3 \pm 0.5$ & $ 7.2 \pm 0.2 $ \\
 {\small NGC7814-DGSAT-3} & $901 \pm 195$ & $ 126.4 \pm 27.1 $ & $-12.9 \pm 0.5$ & $ 7.0 \pm 0.2 $ \\
 {\small NGC7814-DGSAT-4} & $662 \pm 147$ & $ 129.8 \pm 27.9 $ & $-12.3 \pm 0.5$ & $ 6.8 \pm 0.2 $ \\
 {\small NGC7814-DGSAT-5} & $366 \pm 82 $ & $ 165.5 \pm 35.5 $ & $-10.4 \pm 0.5$ & $ 6.0 \pm 0.2 $ \\
 {\small NGC7814-DGSAT-6} & $461 \pm 100 $ & $ 187.6 \pm 40.3 $ & $-12.1 \pm 0.5$ & $ 6.7 \pm 0.2 $ \\
 {\small NGC7814-DGSAT-7} & $1490 \pm 360 $ & $ 257.9 \pm 55.4 $ & $-12.5 \pm 0.5$ & $ 6.9 \pm 0.2 $ \\
  \hline
\end{tabular}
\end{center}
\end{table*}

Finally, it is worth mentioning that the famous Sombrero Galaxy (M~104, distance 
$D$ = 11$\pm$4\,Mpc; Galactic latitude $b^{\rm II}$ = +51\ffcirc1) with its prominent 
bulge is not part of our sample because its inclination is not close enough to edge-on 
(the central core is not hidden by its dusty equatorial disk).  Nevertheless, with an 
inclination of $i$ $\approx$ 85$^{\circ}$ (e.g. Bajaja et al. 1991), it is still close 
to edge-on and the galaxy is located in an ``empty'' field. For studies, which consider 
$i$ $\approx$ 85$^{\circ}$ as a sufficiently rigorous inclination limit, it is thus another 
host galaxy providing an interesting environment worthy of study (see Javanmardi et 
al. 2016). Also interesting may be NGC~4631, the Whale Galaxy ($D$ = 6$\pm$2\,Mpc; $b^{\rm II}$
= +84\ffcirc2), another edge-on galaxy with an equatorial dust lane but a smaller bulge 
(again, we refer to Javanmardi et al. 2016). It is not part of Table~\ref{tab:pairs} because its 
overall shape appears to be asymmetric and distorted due to an interaction with NGC~4656, 
the Hockey Stick Galaxy. The latter is a nearby companion of only slightly lower luminosity 
with a morphology that is also affected by this interaction.

\section{New DGSAT data}

In the following we present deep images of two of the isolated galaxies in our catalogue 
(Table~\ref{tab:singles}): NGC~3669 with a small bulge (Figs.~\ref{fig:ngc3669} and 
\ref{fig:3669_models}) and NGC~7814 with a prominent bulge (Figs.~\ref{fig:ngc7814} and 
\ref{fig:7814_models}), being part of the Dwarf Galaxy Survey with Amateur Telescopes (we refer
to Javanmardi et al. 2016 for an introduction).  Both of these images were obtained with the ROSA 
(POLLUX) 0.4-m Newton f/3.75 telescope containing a CCD camera covering a field of 
$\approx$1.$\!\!^{\circ}$4 $\times$ 1.$\!\!^{\circ}$4 with a pixel size of 1\ffas24 and a full 
width at half maximum resolution of 2\ffas78. Images were obtained with a wide band filter 
known as luminance (Astrodon Gen2 Tru-Balance E-series LRGB) that almost covers the entire 
SDSS $r$ and $g$ bands. NGC~3669 was observed in mid-December 2015 and mid-January 2016 
with a total exposure time of 350\,min, while NGC~7814 was observed at the end of 
September 2016 for 400\,min.

The calibration of the images to SDSS $r$ band, the determination of the limiting surface brightness, 
the search for low surface brightness galaxies and the extraction of their observed parameters was 
carried out in exactly the same way as in Javanmardi et al. (2016). Results referring to individual 
images are explained below.

\subsection{The NGC~3669 (small bulge) field}

The image of the field of NGC~3669 is shown in Figure \ref{fig:ngc3669}. The $5\sigma$ 
limiting surface brightness is 27.7\,mag/arcsec$^2$ and the standard deviation of its 
calibration is 0.02 mag. We have detected eight low surface brightness (LSB) galaxies 
in the field of this image, which are also shown in Figure \ref{fig:ngc3669}. At the distance of
NGC~3669, four of these LSB galaxies are within a projected distance of approximately 200\,kpc to 
NGC~3669, while the other four are off by $\approx$400\,kpc or more. The latter four are (in 
projection) closer to NGC~3625 (which is located near the western edge of our image), and, 
therefore, are named after this galaxy. Seven of these eight LSB galaxies have already been
detected by the SDSS and are catalogued as galaxies, but all of them with low quality 
measurement flags. The remaining one, NGC~3669-DGSAT-3, is a particularly low surface 
brightness galaxy which appears to be very extended. We obtain the observed structural 
parameters of all eight galaxies by modelling them with a Sersic profile using the GALFIT 
software. The results are shown in Table~\ref{tab:results} and Figure~\ref{fig:3669_models}.

\subsection{The NGC~7814 (prominent bulge) field}

Our image of the field of NGC~7814 is shown in Figure \ref{fig:ngc7814}. The 
$5\sigma$ limiting surface brightness is 27.8 mag/arcsec$^2$ and the standard 
deviation of its calibration is 0.01 mag. Seven LSB galaxies were detected in 
this image which are also shown in Figure \ref{fig:ngc7814}. The discovery of 
two of them, namely NGC~7814-DGSAT-1 and 2, has already been reported in the first 
DGSAT paper (Javanmardi et al. 2016). Out of the remaining five, four are 
catalogued by SDSS with low quality measurement flags. Our newly identified 
galaxy, NGC~7814-DGSAT-7, has the lowest surface brightness among our detections. 
Modelling results of all seven galaxies are shown in Table~\ref{tab:results} and 
Figure~\ref{fig:7814_models}.

\begin{figure}
\begin{center}
\includegraphics[scale=0.34]{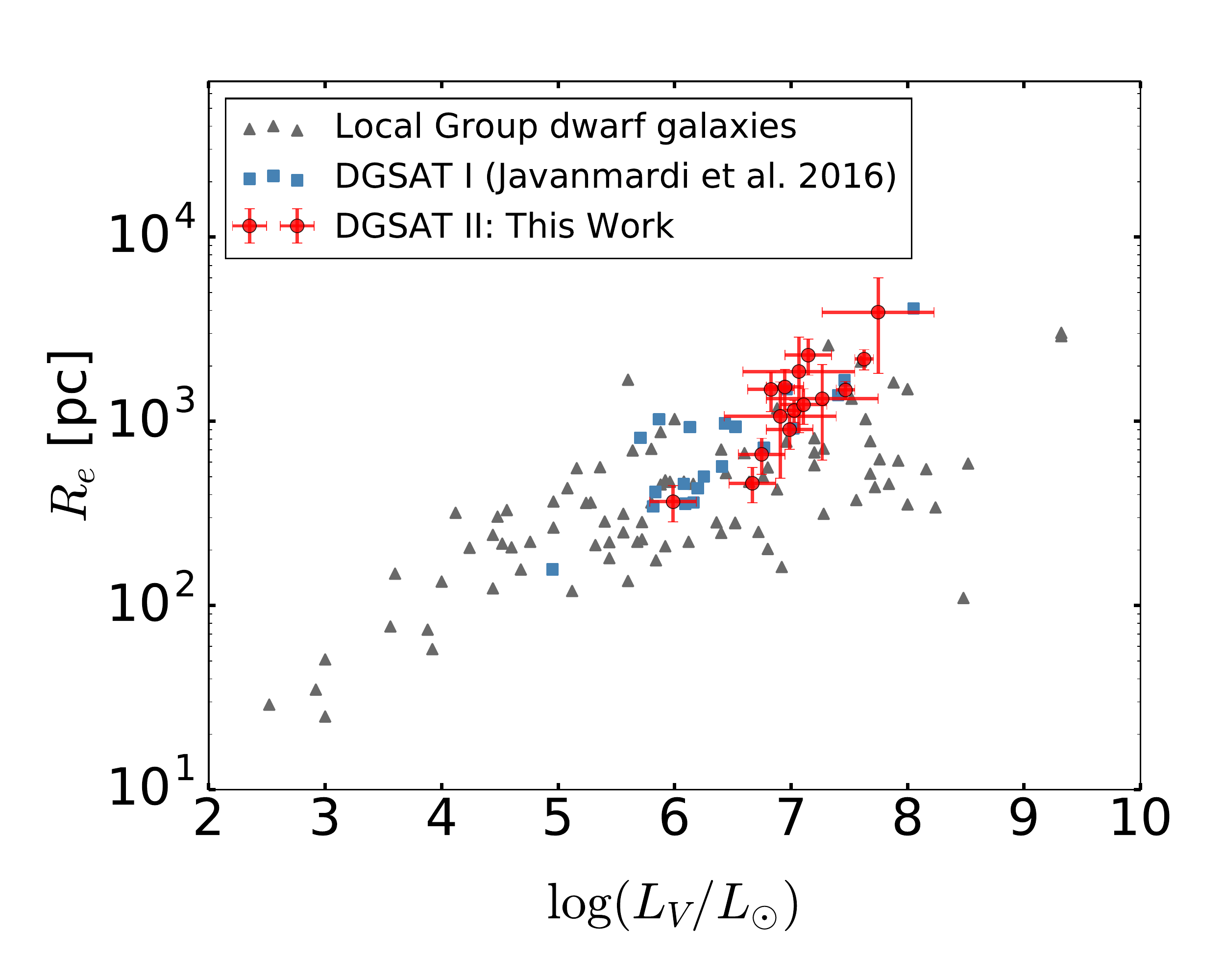}
\includegraphics[scale=0.34]{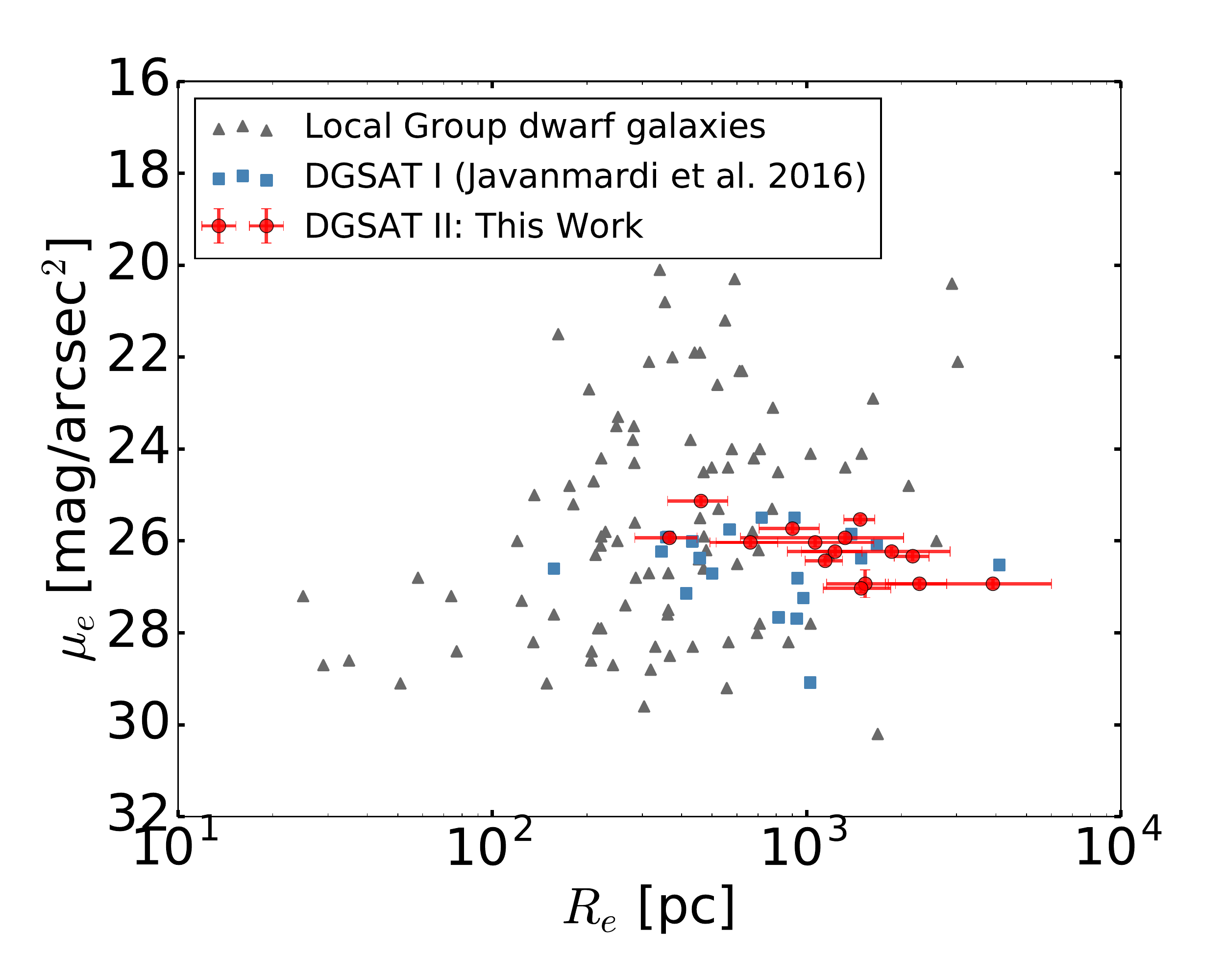}
\includegraphics[scale=0.34]{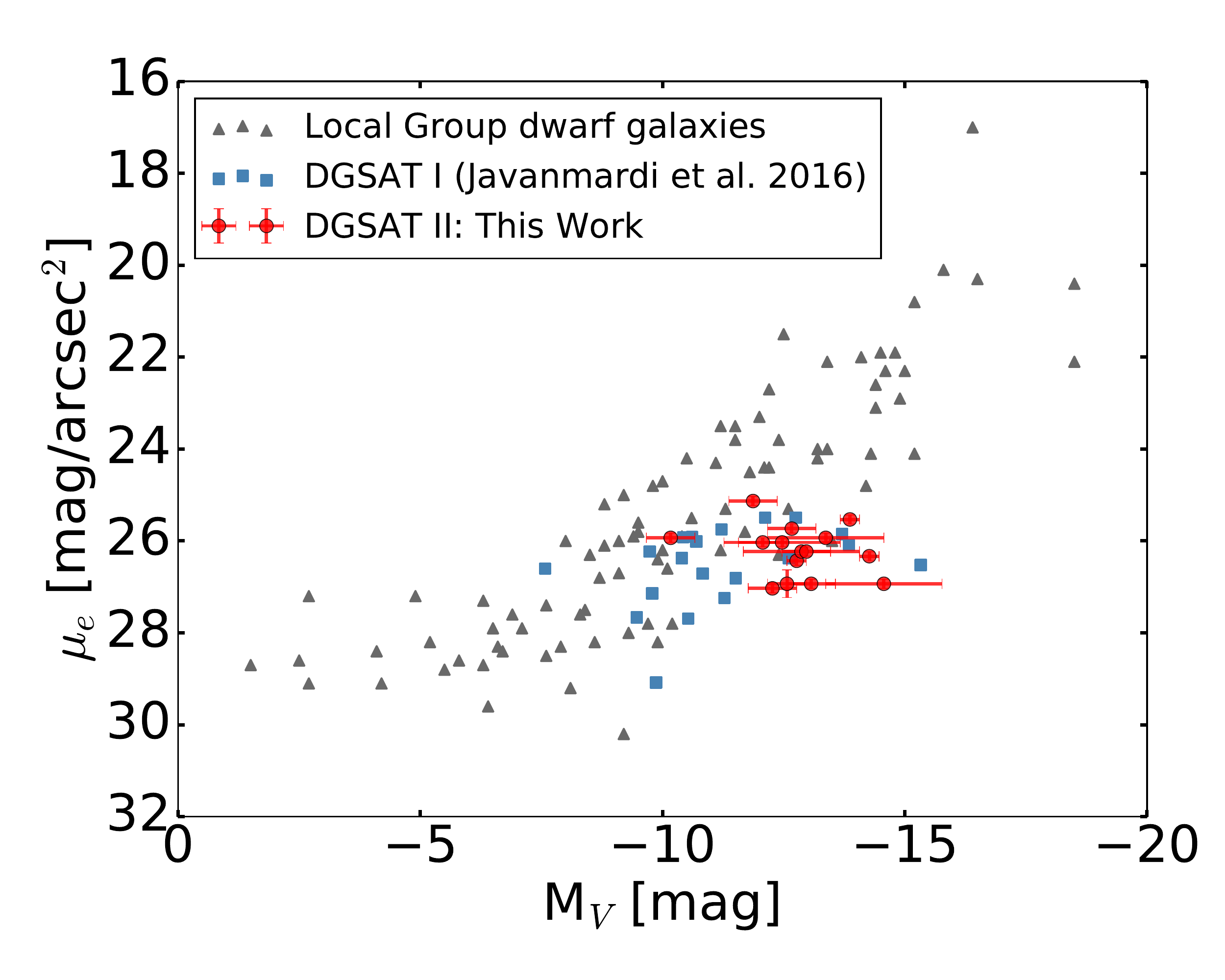}
\caption{A comparison between the properties of the dwarf galaxy candidates 
studied in this work, the first DGSAT results (Javanmardi et al. 2016), and 
the known dwarf galaxies of the Local Group (McConnachie 2012). Top:  
effective radius in pc versus logarithm of V band luminosity in L$_{\odot}$. 
Middle: V band surface brightness in mag/arcsec$^{2}$ versus effective radius 
in pc. Bottom: V band surface brightness in mag/arcsec$^2$ versus absolute 
V band magnitude. The V band quantities were obtained by transforming r 
band magnitudes to V magnitudes using Fukugita et al. (1996). 
\label{fig:withLG}}
\end{center}
\end{figure}

\subsection{Implications}

Our detected objects are characterised by a low surface brightness, 
$\approx$25 -- 27\,mag/arcsec$^2$, cover a range of integrated magnitudes 
from r$_{\rm cal}$ = 17.7 to  20.6\,mag, and exhibit effective radii of 
order 10$''$. Sersic indices are around unity or smaller, consistent 
with what is expected from small galaxies (e.g. Merritt et al. 2014). 
Subtracting the modelled galaxies from the original images (we refer to the 
lowest rows of images in Figs.~\ref{fig:3669_models} and \ref{fig:7814_models}), 
there are no residuals above the noise level. Thus no major irregularities, 
hinting at galaxy interactions, are found. Table~\ref{tab:results_phys} 
provides projected distances of our dwarfs to their putative hosts. Only 
three galaxies show values $<$60\,kpc, with NGC~7814--DGSAT--1 at $\approx$16\,kpc
representing the most critical case. That we see even here no significant distortion 
indicates, that its real distance must be larger. Only one of the targets, 
previously (Javanmardi et al. 2016) detected, NGC~7814--DGSAT--2, shows a 
minor-to-major axis ratio clearly below 0.5. 

Low surface brightness galaxies are difficult targets for follow-up 
measurements, trying to relate them beyond doubt with their major 
host. To provide support for such an association, in addition to their 
close line-of-sight positions, we therefore use the relations already 
mentioned at the end of Sect.\,3. Data are displayed in Fig.~\ref{fig:withLG},
together with the known dwarf galaxies of the Local Group (McConnachie 2012).

Obviously, the galaxies characterised in this article (as those in Javanmardi
et al. 2016) are following the correlations established for the Local Group
galaxies with respect to the connections between luminosity and effective 
radius, effective radius and surface brightness, and absolute magnitude versus
surface brightness. Our galaxies are located at the luminous and extended end 
of these correlations. This is an expected effect since analogues of less 
luminous LG dwarfs are not detectable at the distances to NGC~3669 (43$\pm$23\,Mpc), 
NGC~3625 (37$\pm$4\,Mpc) and NGC~7814 (16$\pm$3)\,Mpc. Therefore associating 
our dwarfs with their putative massive hosts, we find absolute magnitudes of 
$M_{\rm r}$ = --10.4 to --14.8 (Table~\ref{tab:results_phys}). In this context 
we should also note that NGC~3625 and NGC~3669 are located at larger distances 
than the galaxies studied by Karachentsev et al. (2014), Merritt et al. (2014), 
Javanmardi et al. (2016) or M{\"u}ller et al. (2017a,b). Adopting the distances 
to the nearby major galaxies, NGC~3625--DGSAT--4, with a suface brightness much 
fainter than 24\,mag\,arcsec$^{-2}$ and a size $R_{\rm e}$ $>$ 1.5\,kpc, can be 
classified as an ultra-diffuse galaxy (see Merritt et al. 2016). NGC~3669--DGSAT--3, 
our galaxy with the lowest surface brightness, likely also belongs to this class, 
in spite of its highly uncertain effective radius.

A final comment refers to the projected distribution of our LSB galaxies around 
their putative major hosts. In Sects.\,1--3, we outlined potential distributions. 
For our galaxies, no obvious trend is yet apparent. Likely, the number of 
identified dwarfs is still too small (we refer to, for a related case, M{\"u}ller et al. 
2015, 2016, and 2017a and their analyses of the Centaurus group), emphasising the 
urgent need for even more sensitive follow-up studies.

\section{Conclusions}

As shown, there exists a number of promising major edge-on galaxy hosts with well 
determined bulge sizes, where dwarf galaxy distributions can be studied under 
geometrically advantageous conditions. The galaxies in so-called ``empty'' fields
provided by our catalogue (Tables~\ref{tab:singles} and \ref{tab:pairs}) will allow 
for efficient dwarf galaxy surveys of regions with radius $\ga$250\,kpc (1) to test 
whether or not such dwarfs exist at all, (2) to search for the existence of a number of satellites -- 
bulge mass/luminosity relation and (3) to find evidence for anisotropic satellite 
populations. In addition to isolated hosts in such fields, our sample also contains members 
of galaxy pairs, which have the potential to contribute significantly to our understanding 
of more complex galaxy configurations as they are, for example, encountered in our Local Group.  

There are six arguments in support of a close connection of our low surface brightness
(LSB) galaxies with the Milky Way-sized galaxies in their projected neighbourhood:
\begin{itemize}
\item Our targets are, projected onto the plane of the sky, close to their putative
      host galaxies.
\item Our targets are characterized by very low surface brightness.
\item Sersic indices are around unity or smaller.
\item Assuming that our LSB galaxies are at the distance of their supposed main group
      members, their correlation between effective radius and luminosity is consistent with 
      that of Local Group dwarfs, being located at their luminous, extended end. 
\item This also holds for the correlation between effective radius and surface brightness.
\item Finally, absolute magnitudes and surface brightnesses are also consistent with the 
      assumption that our galaxies are located in the vicinity of their major galaxy
      hosts.
\end{itemize}

Concerning the bulge mass versus number of satellites correlation (see Kroupa et al. 2010;
L{\'o}pez-Corredoira \& Kroupa 2016), the results obtained here are four satellite candidates
for NGC~3669 (small bulge) and seven candidates for NGC~7814 (prominent bulge). Right now,
the existence of the relation can neither be excluded nor verified with high confidence
and more observations are mandatory.

\acknowledgements
We are grateful for very good and useful referee comments. BJ would also like to thank the 
Bonn-Cologne Graduate School (BCGS) for Physics and Astronomy for their financial support, 
and also R. Gonzalez-Lopezlira, Z. Shafiee, and Z. Sheikhbahaee for useful discussions on 
image analysis. D.M-D. acknowledges support by the Sonderforschungsbereich (SFB) 881 ``The 
Milky Way System'' of the Deutsche Forschungsgemeinschaft (DFG), in particular through 
subproject A2. The SAO/NASA Astrophysical Data System (ADS) was very helpful when searching 
for relevant articles.

\normalsize
\section*{Appendix}

This appendix provides information on potentially disturbing sources when collecting data on 
dwarf galaxy distributions around major hosts. Adopted distances refer to the average values 
taken from NED, which are displayed for our sources in Col.\,7 of Table~1 and Col.\,8 of Table~2. 
Angular offsets between galaxies are converted into projected distances using the distances ($D$) 
to the edge-on galaxies of our sample. ``Empty field'' indicates a large area free of similarly 
sized galaxies, extending over at least several 100\,kpc.

\smallskip
\smallskip
{\it NGC~7814}: About 1\ffcirc5 to the east, corresponding to a projected distance of 
$\approx$420\,kpc, lies an optically much less prominent galaxy, NGC~14 (distance: 9$\pm$3\,Mpc). 

\smallskip
\smallskip
{\it NGC~100}: Empty field.

\smallskip
\smallskip
{\it UGC~1281}: $\approx$5$^{\circ}$ off M33; NGC~736 and NGC~750 at offsets of $\approx$2$^{\circ}$,
corresponding to $\approx$200\,kpc, in the northwest. Lots of small background galaxies 
clustering most densely $\approx$15$'$ to the east and $\approx$30$'$ to the north.

\smallskip
\smallskip
{\it ESO~0477--G016}: NGC~723 (no specific distance estimate, $V$ = 1485\,km\,s$^{-1}$)
is located $\approx$1$^{\circ}$ in the southwest. This angular distance corresponds to 
$\approx$600\,kpc. Otherwise empty field.

\smallskip
\smallskip
{\it NGC~891}: Two relative to NGC~891 quite inconspicious galaxies, NGC~898 $\approx$25$'$ to the 
south and UGC~1841 $\approx$40$'$ to the north, are with $V$$\approx$5500 and $\approx$6375\,km\,s$^{-1}$ far 
in the background. NGC~906 and NGC~914, $\approx$30$'$ in the southeast and being similarly inconspicious,
have radial velocities of 4680 and 5535\,km\,s$^{-1}$. 

\smallskip
\smallskip
{\it NGC~1886}: No similarly sized galaxy within 2$^{\circ}$ (1\,Mpc).

\smallskip
\smallskip
{\it ESO~121--G006}: The less conspicious galaxy NGC~2205, $\approx$50$'$ to the southeast, is with 
$V$ = 8385\,km\,s$^{-1}$ far in the background. Otherwise empty field.

\smallskip
\smallskip
{\it NGC~2549}: Empty field.

\smallskip
\smallskip
{\it NGC~2654}: There are similarly prominent (w.r.t. angular size) galaxies 2$^{\circ}$--3$^{\circ}$ 
south (NGC~2685) and east (NGC~2742), corresponding to projected distances of 1.0 and 1.5\,Mpc.

\smallskip
\smallskip
{\it UGC~5245}: An optically slightly more prominent galaxy, NGC2974, is located at similar distance
$\approx$2$^{\circ}$ ($\approx$700\,kpc) in the southwest.

\smallskip
\smallskip
{\it NGC~3098}: Empty field.

\smallskip
\smallskip
{\it NGC~3115}: There are two obvious dwarf companions, PGC~29299 and PGC~29300, 5$'$ and 15$'$ -- 20$'$
in the east and southeast, respectively. With respect to galaxies with similar angular extent, empty
field.

\smallskip
\smallskip
{\it UGC~5459}: The less conspicuous galaxy UGC~5460 is located $\approx$75$'$ (projected $\approx$500\,kpc) 
to the south.

\smallskip
\smallskip
{\it NGC~3245A}: The companion galaxy NGC~3245 is located at an offset of $\approx$9$'$ to the
south, corresponding to a projected distance of $\approx$65\,kpc. NGC~3254 ($D$ = 33$\pm$5\,Mpc)
is $\approx$1~degree in the north.

\smallskip
\smallskip
{\it NGC~3669}: Appears to be slightly warped, with the central region edge-on but with 
slightly less inclined outskirts. 40$'$ -- 50$'$ to the south and southeast are NGC~3674 and 
NGC~3683A, 40$'$ -- 50$'$ to the west we find NGC~3613 and NGC~3619 and a few smaller galaxies
at ~30$'$ in the west. These galaxies may be located at a distance of $\approx$30\,Mpc,
which leads to projected distances $\ga$250\,kpc.

\smallskip
\smallskip
{\it UGC~6792}: Low surface brightness (LSB) galaxy PGC37050 $\approx$1$^{\circ}$ (projected 350\,kpc) 
to the south.

\smallskip
\smallskip
{\it NGC~4179}: NGC~4116 and NGC~4123 at similar distance as NGC~4179 (see Table~1) are off 
$\approx$2$^{\circ}$ (projected $\approx$600\,kpc) to the northwest. The background source NGC~4073 
is also seen at an offset of 2~degrees, but in the west.

\smallskip
\smallskip
{\it UGC~7321}: NGC~4204 lies at an offset of $\approx$2$^{\circ}$ ($\approx$600\,kpc) the south.
NGC~4455 is seen at an even larger angular distance. However, both galaxies may be closer to us 
than UGC~7321.

\smallskip
\smallskip
{\it NGC~4244}: The less bright galaxy NGC~4214 is located at similar distance (3$\pm$1\,Mpc) 
$\approx$1\ffcirc5 to the south. At the distance to NGC~4244 this corresponds to a projected linear
offset of 105\,kpc. NGC~4145 and NGC~4151 at distances of 15--20\,Mpc are located $\approx$2$^{\circ}$
in the northwest. 

\smallskip
\smallskip
{\it UGC~7394}: UGC~7370 with a velocity of $V$$\approx$2215\,km\,s$^{-1}$ is located 
$\approx$40$'$ ($\approx$400\,kpc) in the northwest. UGC~7396 with a velocity of 2100\,km\,s$^{-1}$ is 
$\approx$40$'$ to the south. Both galaxies may be seen at a similar distance as UGC~7394.

\smallskip
\smallskip
{\it NGC~4565}: The smaller galaxy NGC~4494 (at similar distance) is located 1\ffcirc15 to the west, 
corresponding to a projected offset of 230\,kpc. NGC~4559 in the north and NGC~4725 in the east,
also at similar distances, show angular offsets twice and three times as large.

\smallskip
\smallskip
{\it NGC~4634}: A pair with NGC~4633, displaced by less than 4$'$ to the north. The pair is surrounded
by Virgo Cluster galaxies ($D$ $\approx$15\,Mpc) at angular offsets $\ga$70$'$ ($\ga$400\,kpc), that is by
M90, M91, NGC~4571, NGC~4639, NGC~4654, NGC~4659, NGC~4689, and NGC~4710.

\smallskip
\smallskip
{\it ESO~575--G061}: There are quite a number of (slightly) less bright galaxies like ESO~575--G060 
in the field. These have radial velocities in excess of 10000\,km\,s$^{-1}$.

\smallskip
\smallskip
{\it NGC5170}: NGC~5247, likely at similar distance, lies almost 2$^\circ$ ($\approx$925\,kpc)
in the east. 

\smallskip
\smallskip
{\it UGC~8880}: Next slightly less prominent galaxies are located about 45$'$ northeast and 55$'$ 
northwest.

\smallskip
\smallskip
{\it NGC~5470}: The less prominent galaxy NGC~5491 is located $\approx$1$^{\circ}$ ($\approx$400\,kpc) 
to the east. Several galaxies (e.g. NGC~5374 (background), NGC~5382 (uncertain distance), and NGC~5387 
(background)) can be found $\approx$2$^{\circ}$ to the west.

\smallskip
\smallskip
{\it UGC~9242}: The galaxy is surrounded by several similarly prominent galaxies at $\approx$1$^{\circ}$
angular distance ($\approx$350\,kpc), i.e. by NGC~5582, UGC~9203, and UGC~9291, which may be located 
at similar or larger distances. 

\smallskip
\smallskip
{\it UGC~9249}: NGC~5665, $\approx$1\ffcirc5 ($\approx$500\,kpc) in the southeast, and NGC~5669 at almost 
2$^{\circ}$ northeast are located at distances similar to that of UGC~9249 (see Table~1).

\smallskip
\smallskip
{\it NGC~5746}: Possibly accompanied by the much smaller galaxy NGC~5740, $\approx$18$'$ to 
the southwest, corresponding to a projected distance of 160\,kpc. Other galaxies (NGC~5690,
NGC~5719, NGC~5774, and NGC~5775) show offsets of at least 2$^{\circ}$ ($\ga$1\,Mpc) and are
located not closer than $\approx$20\,Mpc. 

\smallskip
\smallskip
{\it NGC~5866 and NGC~5907}: Separated by $\approx$1\ffcirc4 (350\,pc). The less conspicious 
galaxies NGC~5905 and NGC~5908, almost a full degree to the south of NGC~5907 and $\approx$1\ffcirc4
east of NGC~5866, belong with $V$$\approx$3350\,km\,s$^{-1}$ to the background. NGC~5879 may be
part of a group composed of NGC~5866 and NGC~5907, but is less bright (location: 
1$^{\circ}$ northwest of NGC~5907 and 1\ffcirc5 north of NGC~5866).

\smallskip
\smallskip
{\it UGC~9977}: UGC~9979, 15$'$ to the south ($\approx$125\,kpc), may be located at 
similar distance but is clearly less prominent. Otherwise empty field. 

\smallskip
\smallskip
{\it ESO~146--G014}: Empty field.

\smallskip
\smallskip
{\it IC~5176}: Empty field.

\smallskip
\smallskip
{\it NGC~7332}: Pair with NGC~7339 in the east (at 5$'$, corresponding
to $\approx$20\,kpc), another galaxy viewed close to edge-on. Otherwise
empty field.

\end{document}